\documentclass[referee,a4paper,12pt,traditabstract]{swsc} 

\usepackage{graphicx}
\usepackage{ulem}
\usepackage{txfonts}
\usepackage{subfigure}
\usepackage{lineno}
\usepackage[authoryear,round]{natbib}
\usepackage[backref]{hyperref}
\usepackage{url}

\hypersetup{colorlinks=true,citecolor=cyan,urlcolor=cyan,linkcolor=blue}

\begin{document}


   \title{Assessing the relationship between spectral solar irradiance and stratospheric ozone using Bayesian inference}


   \author{William T. Ball
          \inst{1}
          \and
          Daniel J. Mortlock\inst{1}\inst{,2}
          \and
          Jack S. Egerton
          \inst{1}
          \and
          Joanna D. Haigh
          \inst{1}
          }

   \institute{Physics Department, Blackett Laboratory, 
				Imperial College London, SW7 2AZ, UK\\
              \email{\href{mailto:william.ball@imperial.ac.uk}{william.ball@imperial.ac.uk}}
         \and
             Department of Mathematics, Imperial College London, SW7 2AZ, UK\\
             }

   \date{Received March 25, 2014; accepted August 11, 2014}

 
  \abstract
{We investigate the relationship between spectral solar irradiance (SSI) and ozone in the tropical upper stratosphere. We find that solar cycle (SC) changes in ozone can be well approximated by considering the ozone response to SSI changes in a small number individual wavelength bands between 176 and $310 \, {\rm nm}$, operating independently of each other. Additionally, we find that the ozone varies approximately linearly with changes in the SSI. Using these facts, we present a Bayesian formalism for inferring SC SSI changes and uncertainties from measured SC ozone profiles. Bayesian inference is a powerful, mathematically self-consistent method of considering both the uncertainties of the data and additional external information to provide the best estimate of parameters being estimated. Using this method, we show that, given measurement uncertainties in both ozone and SSI datasets, it is not currently possible to distinguish between observed or modelled SSI datasets using available estimates of ozone change profiles, although this might be possible by the inclusion of other external constraints. Our methodology has the potential, using wider datasets, to provide better understanding of both variations in SSI and the atmospheric response.}
   \keywords{stratosphere --
                ozone --
                spectral solar irradiance
               }

	 \titlerunning{SSI-Ozone Relationship}
   \maketitle

\section{Introduction}
\label{intro}
The thermal structure and composition of the Earth's upper stratosphere and mesosphere, especially at low latitudes, are determined primarily by the incoming solar irradiance, with the photodissociation of oxygen, nitrogen and water vapour providing the basic constituents for middle atmospheric chemistry. In particular, the decomposition, by ultraviolet (UV) radiation at wavelengths $\lambda <242 \, {\rm nm}$, of molecular oxygen into its component atoms initiates the processes which create ozone, while radiation at wavelengths $\lambda <320 \, {\rm nm}$ decomposes the ozone molecules. The ratio of shorter to longer wavelengths of spectral solar irradiance (SSI) largely determines ozone concentration and the distribution of ozone will respond to changes in that ratio.

The preliminary results from the Spectral Irradiance Monitor (SIM) instrument \citep{HarderFontenla2005a} on the SOlar Radiation and Climate Experiment (SORCE) satellite \citep{Rottman2005}, covering wavelengths between 240 and 2416 nm, suggested a sharper decrease in UV irradiance over the declining phase of solar cycle (SC) 23 \citep{HarderFontenla2009} than had been observed by different instruments over previous SCs \citep{PagaranWeber2011, DelandCebula2012}. None of the subsequent investigations into changes in middle atmosphere composition and tropospheric climate over that period \citep[e.g.,][]{CahalanWen2010,HaighWinning2010,MerkelHarder2011,InesonScaife2011,WangLi2013} have emphatically contradicted the SIM measurements, but questions remain as to their validity.

The difference in UV SC changes, between observations from SORCE and those from prior missions, might suggest that there has been a change in the Sun during the intervening period; for example \cite{HarderFontenla2009} suggested a possible change in the structure of the solar atmosphere. However, other evidence does not indicate a change in the solar surface magnetic structures responsible for UV irradiance variability. For example, the total solar irradiance (TSI) is a constraint on the SSI. Models reconstructing TSI, based-partly on SSI observations prior to SORCE and employing spectral model atmospheres that are time-independent, reproduce TSI observations extremely accurately on all timescales \citep{BallUnruh2012}. UV cycle variability, of the magnitude suggested by SORCE, requires an inverse trend in other parts of the solar spectrum, i.e. the visible and infra-red, in order to be consistent with TSI measurements. Any counterbalance in visible and IR wavelengths to a change in UV cycle trends must exactly compensate so that the TSI remains consistent with reconstructions based on older spectral observations. Solar UV proxies, such as the Mg II index and F10.7 cm radio flux have also not shown any significant change in their behaviour in the last two SCs \citep{Frohlich2009b}. These arguments suggest that either earlier instruments have underestimated UV SC change or the current SORCE instruments are over-estimating it. The latter case is thought to be more likely and to have arisen as a result of insufficient accounting of degradation within the instruments \citep{BallUnruh2011,DelandCebula2012,LeanDeland2012, ErmolliMatthes2013}.

Also on SORCE is the SOLar STellar IrradianCE (SOLSTICE) instrument \citep{McClintockRottman2005}. SOLSTICE covers the wavelength range 115--$320 \, {\rm nm}$, adequate for studies of stratospheric ozone chemistry, although there are uncertainties over its accuracy at wavelengths above 290 nm (personal communication, Marty Snow). These data also show a greater decrease of UV flux over the declining phase of SC 23 than over previous SCs, though not as large as suggested by SIM. Versions of the data have been released with different spectral changes and the implications of these have been investigated in atmospheric models by \cite{BallKrivova2014} and \cite{SwartzStolarski2012}. Both continue to predict the reduction in ozone in the lower mesosphere in response to higher levels of solar activity found by \cite{HaighWinning2010}. However, the estimated amplitude of the ozone response has become smaller with each subsequent data release \citep{BallKrivova2014}. Such an ozone response has not been seen in regression studies of ozone data over the earlier SCs \cite[as, e.g., compiled by][]{AustinTourpali2008}, but was indicated in a preliminary analysis of Sounding of the Atmosphere using Broadband Emission Radiometry (SABER) data over SC 23 by \cite{MerkelHarder2011}. Without any other correlative measurements of solar spectra it is unclear whether the behaviour of SSI (see e.g. \citeauthor{Lockwood2011}, \citeyear{Lockwood2011}) and ozone has been different in recent years, or whether there are errors in the SSI data, the atmospheric models, or in the analysis of the ozone measurements.

The state of understanding of SC SSI changes is currently that they probably lie between the NRLSSI model, at the lower end of SC change, and SORCE observations, at the upper end. With the successor to SORCE/SIM, the Total Solar Irradiance Sensor (TSIS), not expected to launch until at least 2016, and an estimate of the SSI cycle amplitude requiring an accumulation of data over more than half a decade, no further light will be shed on the nature of SC SSI variability for many years. While a thorough examination of the SSI measurement uncertainties should still continue to be undertaken and results revised if necessary, in the meantime, it is imperative that, where possible, other methods that include indirect observations of, or feedbacks from, solar irradiance be employed to better determine SC SSI changes.

The work on the ozone response to different SSI by \cite{HaighWinning2010} has been interpreted by some authors \citep[e.g.,][]{SwartzStolarski2012} as an attempt to ``validate'' the new spectral data, but this is to misunderstand its objectives. Given uncertain measurements of a quantity of interest, it is fundamental to the scientific method to devise tests of the various explanations using whatever other measurements and external information are available. 

A powerful method for implementing such an approach is Bayesian inference, which naturally allows -- indeed, enforces -- the consideration of external information while treating uncertainties in a mathematically self-consistent manner. Our ultimate aim here is to make probabilistic statements about the variability of the UV spectrum from the Sun on the basis of whatever information is available. In the case of parameter estimation, this information comes both from external constraints (which form the priors) and the current data being analysed. \cite{Cox1946} showed that the only self-consistent formalism for manipulating probabilities of this sort is by using Bayes's theorem, and this overall approach is hence known as Bayesian inference. \cite{Jaynes2003} hence described Bayesian inference as `the logic of science' although it is only recently that widespread access to fast computers has made it easy to implement Bayesian methods to problems of practical interest. Bayesian inference has become standard in a number of fields (e.g., cosmology, see \cite{Armitage-CaplanDunkley2011}, and air quality assessment, e.g. \citeauthor{BergamaschiHein2000}, \citeyear{BergamaschiHein2000}). It has also been applied in climate change attribution, e.g. \cite{LeeZwiers2005}. Its utility in studies of the middle atmosphere has long been recognised \citep{BishopHill1984}, and used to good effect \citep{ArnoldMethven2007}, but has considerable further potential.

Here we apply Bayesian inference to the problem of inferring wavelength-dependent changes in SSI from measurements of atmospheric ozone, with a particular focus on how the uncertainties in the ozone measurements impact the SSI inferences. In Section~\ref{datamodels} we describe the atmospheric model, observed stratospheric ozone profiles and the modelled and observed SSI datasets used in this work. We also compare observed profiles with atmospheric model outputs that employ these SSI datasets. A simple linear model and the Bayesian analysis is described in Section~\ref{bayesian}. In Section~\ref{result}, the observed ozone profiles are used to estimate SSI changes. We present our conclusions in Section~\ref{discuss}. All error bars in all figures in this paper are given as one standard deviation. 

\section{Data and models}
\label{datamodels}
\subsection{Atmospheric model}
\label{atmosmodels}

To simulate the atmospheric ozone response to solar irradiance we use a 2D radiative-chemical-transport model, based on \cite{HarwoodPyle1975}, hereafter referred to as the HP model, which has been used to investigate a variety of atmospheric processes from the troposphere to the mesosphere \citep[e.g.,][]{BekkiPyle1996, HarfootBeerling2007, HaighWinning2010}. Time-dependent zonal mean distributions of temperature, momentum and the concentrations of chemical constituents are determined on a grid with 19 latitudes (nearly pole-to-pole, latitude resolution $\pi$/19) and 29 heights ($z$, from the surface to an altitude of $\sim 90 \, {\rm km}$ on a log pressure scale with resolution 0.5 pressure scale heights). The model takes as inputs SSI and monthly mean values of sea-surface temperature (SST) and eddy momentum flux (EMF). The same solar spectrum, which is resolved into 171 wavebands in the wavelength range 116--$730 \, {\rm nm}$, is used to calculate both photodissociation and solar heating rates.

In our investigation we change only the input SSI and, for each case, run the model to (a seasonally varying) equilibrium. We show results for the December solstice, restricting our investigation to equatorial ozone profiles (latitude-weighted 25$^{\circ}$N--25$^{\circ}$S) in the altitude range $\sim30$--$55\, {\rm km}$ (i.e., 18--$0.6 \, {\rm hPa}$). Where there are $N_z = 7$ levels at which ozone concentrations are estimated. The effects of photochemistry on ozone concentrations dominate the influence of transport above 40km; while the opposite is the case below about 25 km \citep{BrasseurSolomon2005}. In the intervening region both play a role. Our model includes a response of the mean circulation to SC irradiance changes, while EMF and SST are fixed. SC variations in the latter two fields have an insignificant effect on ozone in our region of interest: the tropical upper stratosphere.

\subsection{Equatorial ozone profiles}
\label{eqprofiles}
We consider the SC change in ozone, $\Delta O_3(z)$, at each height, $z$, per 100 solar flux units (SFU) of the F$10.7 \, {\rm cm}$ radio flux. 100 SFU represents a change between 2002 and 2008, i.e., approximately the maximum range of variation in SC 23. The cycle changes in F$10.7 \, {\rm cm}$ radio flux scale well with UV cycle changes and, therefore, provide a good proxy. Presenting the stratospheric ozone response in terms of the 100 SFU is a standard in this area of research (e.g. \citeauthor{AustinTourpali2008}, \citeyear{AustinTourpali2008}; \citeauthor{SwartzStolarski2012}, \citeyear{SwartzStolarski2012}) and we follow this approach for simplicity and comparison. The $\Delta O_3(z)$ profiles result from a change in the photolysis rates of O$_2$ and O$_3$. $\Delta O_3(z)$ is, therefore, governed by the change in solar flux, $\Delta F(\lambda)$, with shorter wavelengths generally absorbed at higher altitudes \citep{Meier1991}. Reaction rates are temperature-dependent and this influence on ozone concentration is considered within the model.

\begin{figure}
\noindent\includegraphics[width=40pc, bb=75 370 582 698]{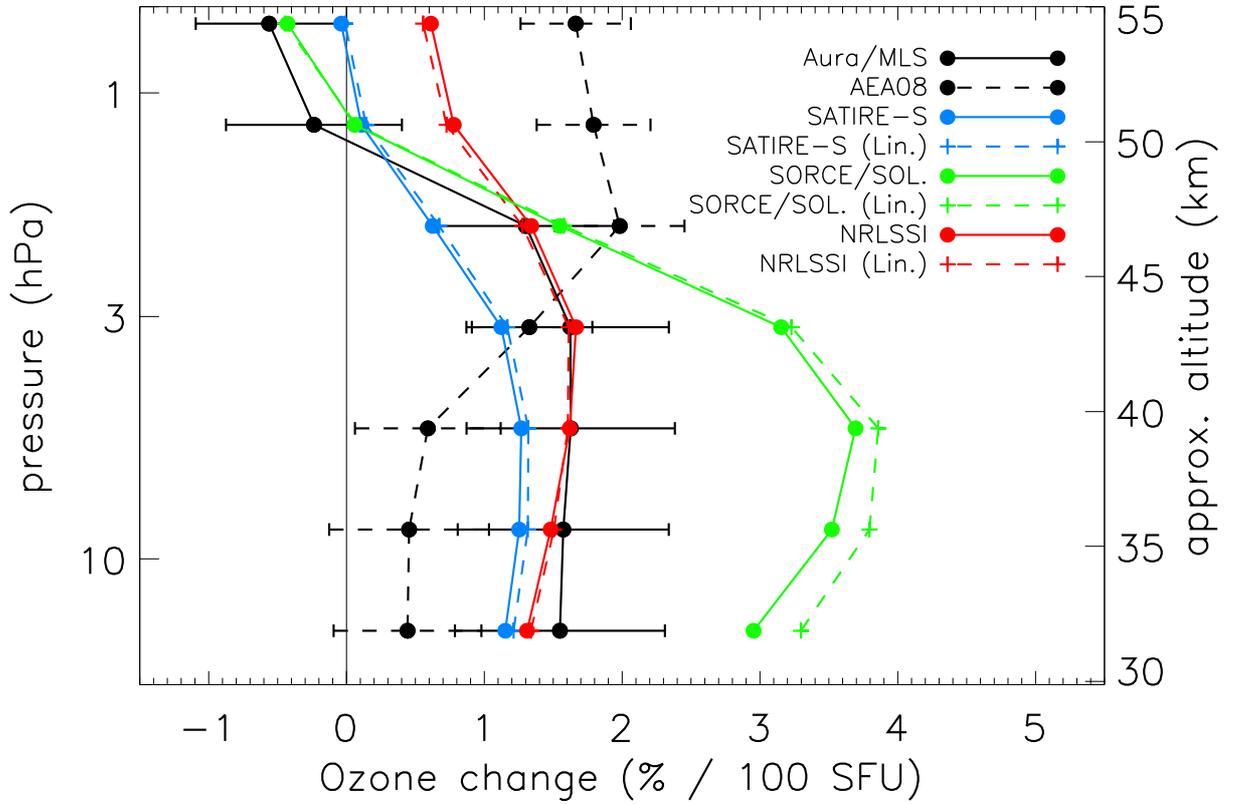}
\caption{The solar cycle signal, $\Delta O_3(z)$, in the profile of equatorial stratospheric: (i) derived from observations, (dashed black curves) \cite{AustinTourpali2008} and (solid black) AURA/MLS; (ii) derived using various solar spectra as input to the HP model: (solid blue) SATIRE-S, (red) NRLSSI and (green) SORCE/SOLSTICE (using SATIRE-S above 290 nm) and (iii) constructed from the (dashed) linear approximation with the SC flux changes given in Table~\ref{ssitable}.}
\label{figsoldataprof}
\end{figure}

We use two observation-based SC equatorial $\Delta O_3(z)$ profiles, both derived using multiple linear regression (MLR) with the F$10.7 \,{\rm cm}$ radio flux as the solar proxy. These are shown in black in Figure~\ref{figsoldataprof}. As with all profiles in this paper, $\Delta O_3(z)$ has been interpolated onto the HP model grid-heights. The dashed black line is from \cite{AustinTourpali2008}, hereafter AEA08, which is the mean of profiles from three different satellite datasets between 1979 and 2003 averaged over $\pm$25$^{\circ}$N, from \cite{SoukharevHood2006}. This profile exhibits an increase in ozone at all altitudes, with a maximum of $\sim$2\% per 100 SFU above $45 \, {\rm km}$ and a minimum of $\sim$0.5\% at $32 \, {\rm km}$. The solid black line is a profile we have derived from Aura/MLS \citep{LayLee2005} ozone data, averaged over $\pm$22.5$^{\circ}$N, for the period August 2004 -- June 2012.  Following the approach of \cite{HaighWinning2010}, we carried out a multiple linear regression analysis of the ozone data with indicators for solar variability, El Ni$\tilde{n}$o-Southern Oscillation and (with two orthogonal indices) Quasi Biennial Oscillation. Although the eight-year period is short relative to the solar cycle, the temporal variation of the F10.7 cm radio flux - with the decaying phase of cycle 23, a minimum near the end of 2008, and the rising phase of cycle 24 - makes it statistically separable from any long-term (linear) trend. The Aura/MLS profile shows a larger positive response than AEA08 below $45 \, {\rm km}$, although, given the uncertainties in these data, the profiles are statistically indistinguishable at these altitudes. At higher levels the behaviour is quite different, with AEA08 showing a larger signal and Aura/MLS a negative change above $50 \, {\rm km}$. The two profiles are derived from different SCs, so the differences may reflect real changes in the UV output from the Sun. However, it should be noted that the Magnesium-II core-to-wing index (see e.g. \citeauthor{SnowMcClintock2005}, \citeyear{SnowMcClintock2005}) and the F$10.7 \,{\rm cm}$ radio flux, both of which are good proxies for the SC UV behaviour, have continued to vary in an expected and consistent way during the recent solar cycles. This may imply that other factors are not properly accounted for in the MLR analyses, or that the statistics are not robust. We make no further judgement on this issue, but take the two profiles as plausible examples with which to demonstrate our technique.

\subsection{Solar spectral irradiance data}
We use two modelled SSI datasets and one observational SSI dataset. The modelled datasets are the Naval Research Laboratory Spectral Solar Irradiance (NRLSSI) \citep{Lean2000, LeanRottman2005} and the Spectral And Total Irradiance REconstruction (SATIRE-S) \citep{FliggeSolanki2000, KrivovaSolanki2003}. SORCE/SIM data have been recalibrated to version 19 (see \citeauthor{BelandHarder2013}, \citeyear{BelandHarder2013}); version 19 is also the first version to extend the time series back to 2003, from 2004. Given that the SORCE/SIM dataset does not extend below 240 nm, we choose to use version 12 of SORCE/SOLSTICE as our observational SSI dataset and consider SORCE/SIM only when comparing the change in flux in wavelength bands above 242 nm. Both SORCE/SOLSTICE and SORCE/SIM instruments have been briefly discussed earlier, so here we complete the descriptions of the solar irradiance datasets by discussing the models NRLSSI and SATIRE-S.

The Spectral And Total Irradiance REconstruction for the Satellite era (SATIRE-S) \citep{FliggeSolanki2000, KrivovaSolanki2003} is a semi-empirical model that assumes all changes in solar irradiance result from the evolution of surface photospheric magnetic flux. Magnetograms and continuum intensity images are used to identify four solar surface components: penumbral and umbral components of sunspots; small-scale bright magnetic features called faculae; and the remaining non-magnetic `quiet' sun. Time-independent spectral intensities are calculated using the FAL-P model atmosphere \citep{FontenlaAvrett1993} modified by \cite{UnruhSolanki1999} for faculae and the spectral synthesis program, ATLAS9 \citep{Kucurz1993}, for the other three components (see \cite{KrivovaSolanki2003} for more details); spectral intensity varies depending on how far the component is from the disk centre and this is taken into account within SATIRE-S. Daily irradiance spectra are then reconstructed by integrating the intensities of the four components as a function of their position on the disk. There is one, main, free parameter in the model relating the magnetic flux detected in a magnetogram to the fraction of that pixel filled with faculae; this free parameter is fixed to an observational time series, usually TSI (see \cite{BallUnruh2012} for further details).

In the NRLSSI model \citep{Lean2000, LeanRottman2005}, spectral irradiance is calculated empirically from observations. The evolution of magnetic flux in the form of sunspots and faculae are calculated using the disk-integrated Mg II and photospheric sunspot indices (PSI). Below 400 nm, multiple regression analysis is performed with detrended (i.e. rotational) UARS/SOLSTICE observations to determined the spectral irradiances; detrended data are used to prevent effects from instrument degradation being introduced to cycle or longer trends. Above 400 nm, spectra are calculated using the models by \cite{SolankiUnruh1998}, with the facular and sunspot contrasts scaled to agree with TSI observations on solar cycle time scales.

The relative SC percentage changes per 100 SFU of SATIRE-S for six UV bands are given in column 3 of Table~\ref{ssitable}. The change of 100 SFU represents the observed change in the solar F10.7 cm radio flux that is approximately the change between solar maximum in 2002 and solar minimum in 2008. The SC change for NRLSSI and SORCE/SOLSTICE in multiples of SATIRE-S are given in columns 4 and 5, respectively. Additionally, and for reference, we include the SC of SORCE/SIM relative to SATIRE-S in column 6, which can only be given for bands between 242 and 310 nm.

\begin{table*}
\caption{SC flux changes within six wavelength bands per 100 SFU. The \% change of SATIRE-S is given in column~(3). Changes in terms of multiples of SATIRE-S are given for NRLSSI, SORCE/SOLSTICE and SORCE/SIM are given in columns~(4) -- (6). SORCE/SIM v19 covers wavelengths from 240 nm, so only bands k = 4--6 are given.}
\begin{center}
\small
\begin{tabular}{cccccc}
\hline 
(1) & (2) & (3) & (4) & (5) & (6) \\
$\lambda_{\rm min}$-$\lambda_{\rm max}$ & Index & SATIRE-S & NRLSSI & SORCE/ & SORCE/ \\
 & $k$ & & & SOLSTICE & SIM \\
 nm & & \%/100SFU & /SAT & /SAT & /SAT \\
\hline
176-193 & 1 & 6.1 & 1.2 & 1.5 & - \\
193-205 & 2 & 4.9 & 1.3 & 2.3 & - \\
205-242 & 3 & 2.7 & 1.0 & 3.2 & - \\
242-260 & 4 & 2.4 & 0.8 & 2.3 & 3.8 \\
260-290 & 5 & 1.2 & 0.7 & 3.0 & 4.4 \\
290-310 & 6 & 0.4 & 0.6 & - & 5.8 \\
\hline 
\end{tabular}
\end{center}
\label{ssitable}
\end{table*}

The stated uncertainty range of 0.5\% per annum in SORCE/SOLSTICE data implies a maximum uncertainty of $\sim$3.4\% over the period considered at the 1$\sigma$ level. We do not use SORCE/SOLSTICE in band 6, between 290 and 310 nm due to this band not being considered reliable (personal communication, Marty Snow); instead we use SATIRE-S at these wavelengths and we consider this reasonable for demonstrating the different ozone responses to the SSI observations and models. We do not quote the uncertainty range for SORCE/SOLSTICE in Table~\ref{ssitable}, but note that it encompasses SATIRE-S in all bands, except between 193 and 242 nm, and NRLSSI in all bands outside of 193 and 260 nm. We do, however, plot the uncertainty range in Fig.~\ref{figmls} as a Gaussian (green lines).

The $\Delta O_3(z)$ profiles estimated using the HP model with the different SSI datasets are shown in Figure~\ref{figsoldataprof} (colored solid lines). SORCE/SOLSTICE data are not available prior to May 2003. The change in the F10.7 cm radio flux between the 81-day average periods centred on 2003 June 15 and 2008 December 15 is $\sim$67 SFU so we rescale the spectra in all the datasets to the 81 day period centred on 2002 October 21, representing a change of 100 SFU since 2008 December 15, with a scaling factor of 1.629. We treat all datasets in the same way -- this scaling is incorporated into the results listed in Table~\ref{ssitable} and is justified as SC changes in UV scale well with the F10.7 cm radio flux. We find this scaling shows good agreement with the cycle change in UV flux for the 2002 to 2008 time period encompassing a full 100 SFU change in SATIRE-S and NRLSSI in the declining phase of SC 23. In table~\ref{f107ratio} we give the percentage change of SATIRE-S and NRLSSI for 2002--2008 and 2003--2008 (columns 3 \& 4 and 6 \& 7, respectively) and the ratio of the former period with the scaled (by 1.629) latter period. Even though both models construct cycle changes differently, they both give ratios close to 1.00, justifying the use of the F10.7 cm radio flux in scaling up the SORCE/SOLSTICE fluxes to 100 SFU. The HP model requires solar fluxes up to 730 nm, so for SORCE/SOLSTICE runs we use SATIRE-S fluxes above 290 nm.

\begin{table*}
\caption{The percentage change in flux in each wavelength band of SATIRE-S (abbreviated as SAT) and NRLSSI (NRL) between October 2002 (abbr. 02) and December 2008 (abbr. 08) (columns~(3) and (6)) and June 2003 (abbr. 03) and December 2008 (columns~(4) and (7)). The scaling factor to estimate the solar cycle change per 100 SFU in 2002, relative to 2008, based on the F10.7 cm radio flux in 2002, 2003 and 2008, is 1.629. The ratio of the rescaled fluxes to 2002 from between 2003 and 2008, in each band compared to the actual fluxes reconstructed by the models between 2002 and 2003 are given for SATIRE-S and NRLSSI in columns~(5) and (8), respectively.}
\begin{center}
\small
\begin{tabular}{cccccccc}
\hline 
(1) & (2) & (3) & (4) & (5) & (6) & (7) & (8) \\
$\lambda_{\rm min}$-$\lambda_{\rm max}$ & Index & SAT & SAT & SAT & NRL & NRL & NRL \\
 & $k$ & [02--08] & [03--08] & 1.629*[03--08] & [02--08] & [03--08] & 1.629*[03--08] \\
 & $ $ & /[08]*100 & /[08]*100 & /[02--08] & /[08]*100 & /[08]*100 & /[02--08] \\
 nm & & \% & \% & & \% & \% & \\
\hline
176-193 & 1 & 6.0 & 3.8 & 1.03 & 7.7 & 4.7 & 0.99 \\
193-205 & 2 & 4.7 & 3.0 & 1.03 & 6.2 & 3.8 & 0.99 \\
205-242 & 3 & 2.6 & 1.7 & 1.03 & 2.8 & 1.7 & 0.99 \\
242-260 & 4 & 2.3 & 1.5 & 1.03 & 2.0 & 1.2 & 0.99 \\
260-290 & 5 & 1.2 & 0.8 & 1.03 & 0.8 & 0.5 & 1.00 \\
290-310 & 6 & 0.4 & 0.3 & 1.03 & 0.3 & 0.2 & 1.01 \\
\hline 
\end{tabular}
\end{center}
\label{f107ratio}
\end{table*}

In Fig~\ref{figsoldataprof}, it can be seen  that the different SC spectral changes give rise to very different $\Delta O_3(z)$ profiles. The SATIRE-S profile (blue) lies within the 1$\sigma$ error bars of Aura/MLS. The NRLSSI profile (red) is in good agreement with Aura/MLS below $50 \, {\rm km}$, but does not show the negative response at higher levels. The NRLSSI profile shown here is also very similar to that presented in Fig.~4 of \cite{SwartzStolarski2012} using the GEOSCCM 2D model. The large changes in the SORCE/SOLSTICE UV data produce much larger ozone changes through most of the stratosphere but with a profile shape (green) similar to that produced by the SATIRE-S spectrum and in the MLS analysis. None of the modelled profiles compare well with the AEA08 profile.

It is important to highlight that the magnitude of the mesospheric response to SORCE solar flux depends on the version used; the SORCE/SOLSTICE version 12 data used here is the second recalibration since the SORCE/SOLSTICE data used by \cite{HaighWinning2010}. The latter was a hybrid of SORCE/SIM version 17 and SORCE/SOLSTICE version 10. \cite{BallKrivova2014}, using the same model set up as in this paper and in \cite{HaighWinning2010}, showed that using just SORCE/SOLSTICE version 10 data below 310 nm resulted in a slightly larger negative response of -1.6\% in ozone at 55 km compared to the -1.2\% in \cite{HaighWinning2010}. The use of version 12 SORCE/SOLSTICE data sees this negative response reduce to -0.2\%. Thus, results using the older SORCE/SOLSTICE should not be considered reliable. Even though the latest version of SORCE/SOLSTICE data should be considered in future studies, there is still a large range of SC change estimates encompassed by SORCE/SOLSTICE and NRLSSI, so a large uncertainty remains in our knowledge in SSI SC changes. In what follows we propose a method for limiting this range based on ozone observations.

The results presented in Figure~\ref{figsoldataprof} and discussed above might suggest, based on the Aura/MLS profile, that SATIRE-S provides a good representation of SC flux changes. But, in the following we show that none of the SSI datasets can (generally) be said to be more representative of the true behaviour of the Sun over the SC than the others, despite large differences in cycle variability and in the ozone response.

\subsection{Linear approximation}
\label{linearity}
The inference problem that will follow is greatly simplified by the fact that, at least over the range of physically plausible values (see Section~\ref{priors}), $\Delta O_3(z)$ varies approximately linearly with changes in solar flux in broad spectral bands. This result, as will be defined in detail in the following, allows a simple \textit{linear} model approximation to the HP model to be set up and for the expected SSI input to be calculated by inversion for any given ozone profile.

We linearise around a reference ozone change profile, $\Delta O_3^{\rm ref}(z)$, chosen to minimise the deviation of the linear approximation from the HP model results (see Section~\ref{bands}), and scale (for convenience, arbitrarily) the change in SSI by that of SATIRE-S. We then find that the total ozone response to changes across the entire spectrum can be accurately reproduced as a sum of its responses in a limited number, $N_k$, of spectral bands leading to
\begin{equation}
\label{equation:linear}
\Delta O_3(z) - \Delta O_3^{\rm ref}(z) = \sum_{k = 1}^{N_k} M_{z,k}\frac{\Delta F_{k}}{\Delta F_{k}^{\rm SAT}} .
\end{equation}
where $M_{z,k} = \Delta O_3^{\rm SAT}(z) - \Delta O_3^{\rm ref}(z)$ is the response to SATIRE-S changes in waveband $k$ (see Section~\ref{bands} for an explanation of how $\Delta O_3^{\rm ref}(z)$ and $M_{z,k}$ are constructed), $\Delta F_{k}$ is the SC change in flux in band k and $\Delta F_{k}^{SAT}$ is the SC change in SATIRE-S in band k, the ratio of which are the parameters $\mathbf{f}$.

\subsection{Choice of spectral bands}
\label{bands}
\cite{Ball2012Thesis} showed that cycle changes in spectral regions below 176 nm and above 310 nm have an insignificant effect on the $\Delta O_3(z)$ profile at the heights we consider in this study. Six bands were selected in this study, based on: (i) the approximate wavelength at which O$_{2}$ or O$_{3}$ photolysis rates decline significantly, e.g., at 242 nm; and (ii) the wavelengths in SSI data at which sudden jumps are seen in the variability of SC flux.

We restrict $N_k$ to six for two reasons: (i) tests show this number provides profile accuracy well within the limits imposed by other parameters in the study and it becomes increasingly costly to test the linear model as $N_k$ increases; and (ii) in general the wavelength response of adjacent narrow bands have similar $\Delta O_3(z)$ profile shapes leading to a redundancy (except at a boundary wavelength for a photochemical reaction, as is the case at 242 nm).

\begin{figure}
\noindent\includegraphics[width=40pc, bb=95 370 577 692]{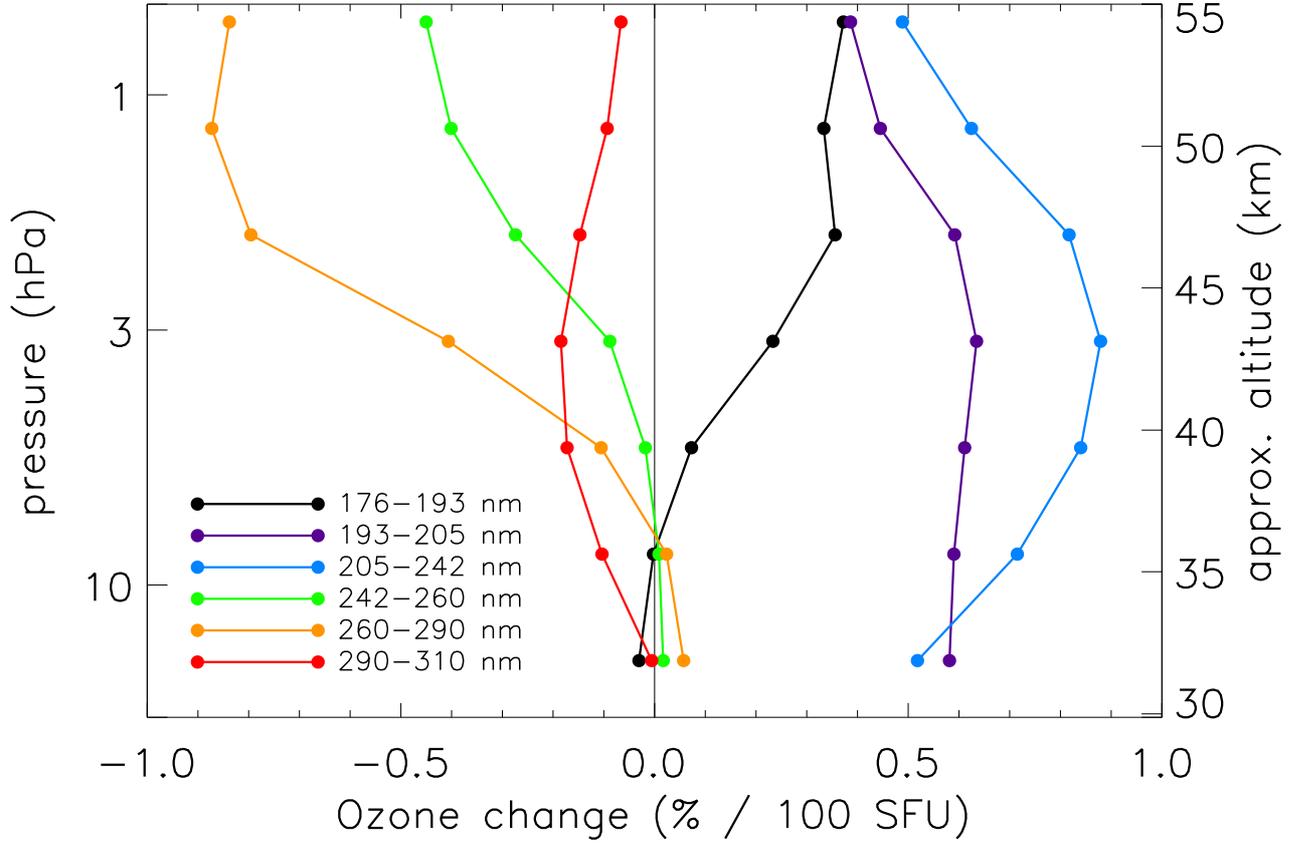}
\caption{The equatorial $\Delta O_3(z)$ profiles resulting from the change in SATIRE-S spectrum separately in the six spectral bands used in the linear model. These are the columns of matrix \textbf{M} defined in the section~\ref{bayesian}.}
\label{figwvlbands}
\end{figure}

The six bands are bounded at 176, 193, 205, 242, 260, 290 and 310 nm. Outside 176--310 nm, we use only the prescribed SC flux change from SATIRE-S. In order to fit $\Delta O_3^{\rm ref}(z)$ and the set of $M_{z,k}$ values, 1000 random uniformly sampled $\Delta O_3(z)$ profiles were produced using the HP model using a plausible range of possible SC changes (see Section~\ref{priors}). We construct the linear profiles that make up each band encoded within $M_{z,k}$, and the reference ozone profile $\Delta O_3^{\rm ref}(z)$, by using Monte Carlo methods to find the values in both $\Delta O_3^{\rm ref}(z)$ and $M_{z,k}$ that minimises, using least squares, the difference between the linear model given in Equation~\ref{equation:linear} and the ozone profiles numerically calculated using the HP model within the prior space. The $\Delta O_3(z)$ profiles due to a SC change in SATIRE-S for each band as given in Table~\ref{ssitable}, are shown in Figure~\ref{figwvlbands}; the shape of these profiles show similarities to those presented in Figure~9 of \cite{SwartzStolarski2012} and Figure~2 of \cite{ShapiroRozanov2013}, although both studies consider different bands (and models) to those presented here. 

To test the assumption of linearity we compare the linear fit described above to 1000 random HP model runs giving $N_s = 7000$ points of comparison, since there are $N_z = 7$ altitude points of comparison in each of the 1000 runs. In general the linear approximation is good relative to the uncertainties in Aura/MLS and AEA08 profiles. The smallest AEA08 1$\sigma$ error from the $N_z = 7$ altitude bins is 0.40\%; it is 0.53\% for Aura/MLS. Over 98\% of the $N_s$ points deviated by less than 0.04\% from the HP model, i.e., an order of magnitude smaller deviation than the smallest error, while the largest deviation was less than 0.1\%, a quarter the smallest 1$\sigma$ error. Hence, we can be confident that adopting this linear approximation to the HP model will not significantly affect the parameter inferences.

We show by example, in Figure \ref{figsoldataprof}, that the linear model (dashed coloured curves) can reproduce well the SATIRE-S (blue), NRLSSI (red) and SORCE/SOLSTICE (green) profiles when the flux changes as given in Table~\ref{ssitable} are applied. Although there are small differences, the agreement is excellent. Therefore, the linear model can be used to approximate the HP model and that, due to the simplicity of its construction, this can lead one to easily determine the SSI flux changes in each band that produces any (reasonable) ozone profile. However, ozone profiles can be constructed using a range of SSI flux changes and still remain within the ozone profiles uncertainties: this is a new result that makes it far more straightforward to constrain SSI from ozone measurements. We do that using the formalism of Bayesian inference.

\section{Bayesian inference of SSI from ozone profiles}
\label{bayesian}

Given a measured ozone profile and the associated measurement uncertainties, Bayesian inference can be used to obtain the tightest reasonable constraints on the SSI. The data, $\mathbf{d} = (d_1, d_2, \ldots, d_{N_z})^\mathrm{T}$, are the measured values of $\Delta O_3(z)$, relative to the reference profile, in each of the $N_z = 7$ height bins defined in Section~\ref{atmosmodels}. The model parameters to be constrained are the factors, $\mathbf{f} = (f_1, f_2, \ldots, f_{N_k} )^\mathrm{T}$, by which the SC change in flux is varied in each of the $N_k = 6$ wavelength bands defined in Section~\ref{bands} and Table~\ref{ssitable}. The aim is to calculate the posterior distribution in the model parameters, which is given (up to an unimportant normalisation constant) by Bayes's theorem as 

\begin{equation}
\label{bth}
\mathrm{Pr}(\mathbf{f} | \mathbf{d})
  \propto \mathrm{Pr}(\mathbf{f}) \, \mathrm{Pr}(\mathbf{d} | \mathbf{f}),
\end{equation}
where $\mathrm{Pr}(\mathbf{f})$ is the prior on the model parameters (which encodes any additional external information) and $\mathrm{Pr}(\mathbf{d} | \mathbf{f}$) is the likelihood (which encodes information contained in the measurements). 

In the case of a parameter estimation problem like that considered here, the full result of Bayesian inference is the posterior probability distribution in the model parameters.  This distribution can then be used to calculate summary statistics, such as credible intervals, estimates and errors, or most probable models, but it is the distribution itself which is the final answer. It is important to note that it is the (posterior) probability which is distributed over the parameter space -- our final state of knowledge is imperfect, an inevitable consequence of the noisy data and prior uncertainties about the model.

We now go through the definition of the model and the steps required to evaluate (an approximation to) this posterior distribution.

\subsection{Parameter priors}
\label{priors}
Some constraints can be placed on $\mathbf{f}$ even without considering the measured ozone profiles; these are encoded in the prior distribution $\mathrm{Pr}(\mathbf{f})$. The priors presented below allow for a SC UV change that exceeds all observations and model estimates. We use these as very conservative constraints reflecting the large uncertainty in the current state of knowledge of SC UV changes. We have performed experiments using more restrictive priors (not shown) and find that the posterior results are not too sensitive to the exact choice (and combination) of the priors.

At the longer wavelengths, our priors restrict the allowed uncertainty on the SORCE/SOLSTICE observations, i.e. of 3.4\%. While these uncertainties are very large compared to nominal estimates of SC changes, we use them as outside possibilities given the available observational data.

The adopted prior constraints are that the: 
\begin{enumerate}
\item maximum change in all bands is limited to 6 $\times$ SATIRE-S (which encompasses all the SSI datasets and is 50\% larger than the largest change in SORCE/SOLSTICE;
\item minimum change in bands $k=1$--3 is zero, as all observations and models below 242 nm have positive SC changes;
\item minimum change in bands $k=4$--6 is $-1$; this negative lower limit on the prior is set as the uncertainty of the SC changes at these longer wavelengths does not necessarily exclude negative changes; by this prior, the SORCE/SOLSTICE estimate of $-2.7$ in band 6 is rejected a priori since, as stated in section~\ref{eqprofiles}, this negative solar cycle change should not be considered reliable;
\item bands $k=1$--3 have higher relative SC changes than the most variable of bands $k=4$--6 (reflecting the same trend seen in the datasets considered here). 
\end{enumerate}

We summarise our adopted priors 1, 2 and 3 in Table~\ref{priorstable}. The priors are enforced by applying rejection sampling, as described below, by rejecting candidate samples if each randomly generated set of $\mathbf{f}$ values, or model, do not satisfy the criteria of each prior. All models which satisfy these constraints are considered to be equally plausible a priori, implying a $\mathrm{Pr}(\mathbf{f}$) is uniform within this somewhat complicated region of parameter space.

\begin{table*}
\caption{The minimum and maximum allowed values of the priors, in multiples of SATIRE-S SC changes, are given in columns 3 and 4, respectively, for each of the wavelength bands and their indices given in columns 1 and 2, respectively.}
\begin{center}
\small
\begin{tabular}{cccc}
\hline 
(1) & (2) & (3) & (4) \\
$\lambda_{\rm min}$-$\lambda_{\rm max}$ & Index & Prior min. & Prior max. \\
 nm & & /SAT & /SAT \\
\hline
176-193 & 1 & 0 & 6 \\
193-205 & 2 & 0 & 6 \\
205-242 & 3 & 0 & 6 \\
242-260 & 4 & -1 & 6 \\
260-290 & 5 & -1 & 6 \\
290-310 & 6 & -1 & 6 \\
\hline 
\end{tabular}
\end{center}
\label{priorstable}
\end{table*}

\subsection{Likelihood}
\label{likeli}
Under the linear approximation described in Section~\ref{linearity}, the data are related to the model parameters by

\begin{equation}
\textbf{d} = \textbf{M}\textbf{f} + \textbf{n},
\end{equation}
where \textbf{M} is the $N_z \times N_k$ response matrix containing all values of $M_{z,k}$ (whose columns are the six $\Delta O_3(z)$ profiles in Figure~\ref{figwvlbands}) and \textbf{n} is the measurement noise of the ozone profile. The noise is assumed to be additive and normally distributed, and so its statistical properties are entirely characterised by the noise covariance matrix, $\textbf{N} = \langle\textbf{n}\textbf{n}^{\mathrm{T}}\rangle$, where the angle brackets denote an average over noise realisations. Here the noise in different bins is taken to be independent, but could in the future be adjusted to incorporate the correlations induced by the MLR reconstruction.

The combined assumptions of linearity and Gaussianity imply that the likelihood is given by
\begin{equation}
\mathrm{Pr}(\textbf{d}{|}\textbf{f})
\!=\!\frac{1}{|(2\pi)^{N_z}\!\textbf{N}|^{\!1\!/\!2}}
\!\mathrm{exp}\!\left[\!-\!\frac{1}{2}\!(\textbf{d}\!-\! \textbf{M} \textbf{f})^{\mathrm{T}}\!\textbf{N}^{-1}\!(\!\textbf{d}\!-\!\textbf{M}\textbf{f})\!\right]\!.
\end{equation}

\subsection{The posterior distribution and sampling}
\label{sampling}

Provided the matrix $\mathbf{C} = \textbf{M}^{\mathrm{T}}\textbf{N}^{-1}\textbf{M}$ is non-singular (as is the case here), the unique maximum likelihood model is 

\begin{equation}
\label{eq2}
\textbf{f}_{\mathrm{ML}} 
  = (\textbf{M}^{\mathrm{T}}\textbf{N}^{-1}\textbf{M})^{-1}\textbf{M}^{\mathrm{T}}\textbf{N}^{-1}\textbf{d}
  = \mathbf{C}^{-1} \textbf{M}^{\mathrm{T}}\textbf{N}^{-1}\textbf{d}.
\end{equation}

\noindent From Equation~\ref{bth}, the posterior is then given (again, up to a normalising constant) by 

\begin{equation}
\mathrm{Pr}(\textbf{f}{|}\textbf{d}) \propto \mathrm{Pr}(\textbf{f}) \, \mathrm{exp}\left[-\frac{1}{2}(\textbf{f}-\textbf{f}_{\mathrm{ML}})^{\mathrm{T}}\textbf{C}^{-1}(\textbf{f}-\textbf{f}_{\mathrm{ML}})\right].
\label{posterior}
\end{equation}
The constraints on any coefficient $f_k$ are given by integrating $\mathrm{Pr}(\textbf{f}{|}\textbf{d})$ over the other $N_k - 1$ coefficients to obtain the marginal distribution $\mathrm{Pr}(f_k|\textbf{d})$.

The restrictions on the model parameters placed by the priors mean that the posterior is significantly more complicated than if it were just a multivariate normal distribution, and so all derived quantities (fitted values, uncertainties, etc.) were calculated by generating samples from $\mathrm{Pr}(\textbf{f}{|}\textbf{d})$. This was achieved using rejection sampling, with two distinct steps: first samples were drawn from an envelope density given by a multivariate normal of mean $f_{\rm ML}$ and covariance $C$; then only those samples which satisfied the prior criteria listed in Section~\ref{priors} were retained. For all the results presented below, $10^{6}$ samples were drawn from the relevant posterior; parameter estimates and errors were then calculated directly from the samples.

\section{Results}
\label{result}

\begin{table*}
\caption{SC flux changes within the six wavelength bands per 100 SFU given in terms of multiples of SATIRE-S. The maximum likelihood SSI SC change without priors of Aura/MLS and AEA08 are given in columns~(3) and (5), respectively. The maximum likelihood SC SSI changes with priors inferred for the Aura/MLS and AEA08 $\Delta O_3(z)$ profiles are given in columns~(4) and (6), respectively; the inferred changes for halved errors on the $\Delta O_3(z)$ profiles are given in brackets.}
\begin{center}
\small
\begin{tabular}{ccccccccc}
\hline 
(1) & (2) & (3) & (4) & (5) & (6) \\
$\lambda_{\rm min}$-$\lambda_{\rm max}$ & Index & Aura/MLS & Aura/MLS est. & AEA08 & AEA08 est. \\
 & $k$ & $f_{\mathrm{ML}}$ & $\sigma$=1 ($\sigma$=0.5) & $f_{\mathrm{ML}}$ & $\sigma$=1 ($\sigma$=0.5) \\
 nm & & /SAT & /SAT & /SAT & /SAT \\
\hline
176-193 & 1 & 6.2 & 2.4 (4.5) & 3.0 & 2.2 (2.7) \\
193-205 & 2 & -2.7 & 0.9 (0.9) & -1.4 & 0.4 (0.4) \\
205-242 & 3 & 6.1 & 1.7 (1.6) & 2.7 & 0.6 (0.6) \\
242-260 & 4 & 7.8 & 1.8 (1.9) & 2.1 & 0.6 (0.6) \\
260-290 & 5 & 0.6 & 1.8 (2.7) & -1.3 & -1.0 (-0.8) \\
290-310 & 6 & 13.3 & 0.9 (1.3) & 6.7 & 2.2 (2.4) \\
\hline 
\end{tabular}
\end{center}
\label{fvaltable}
\end{table*}

\begin{figure*}
\noindent\includegraphics[width=25pc,angle=90, bb=127 118 512 724]{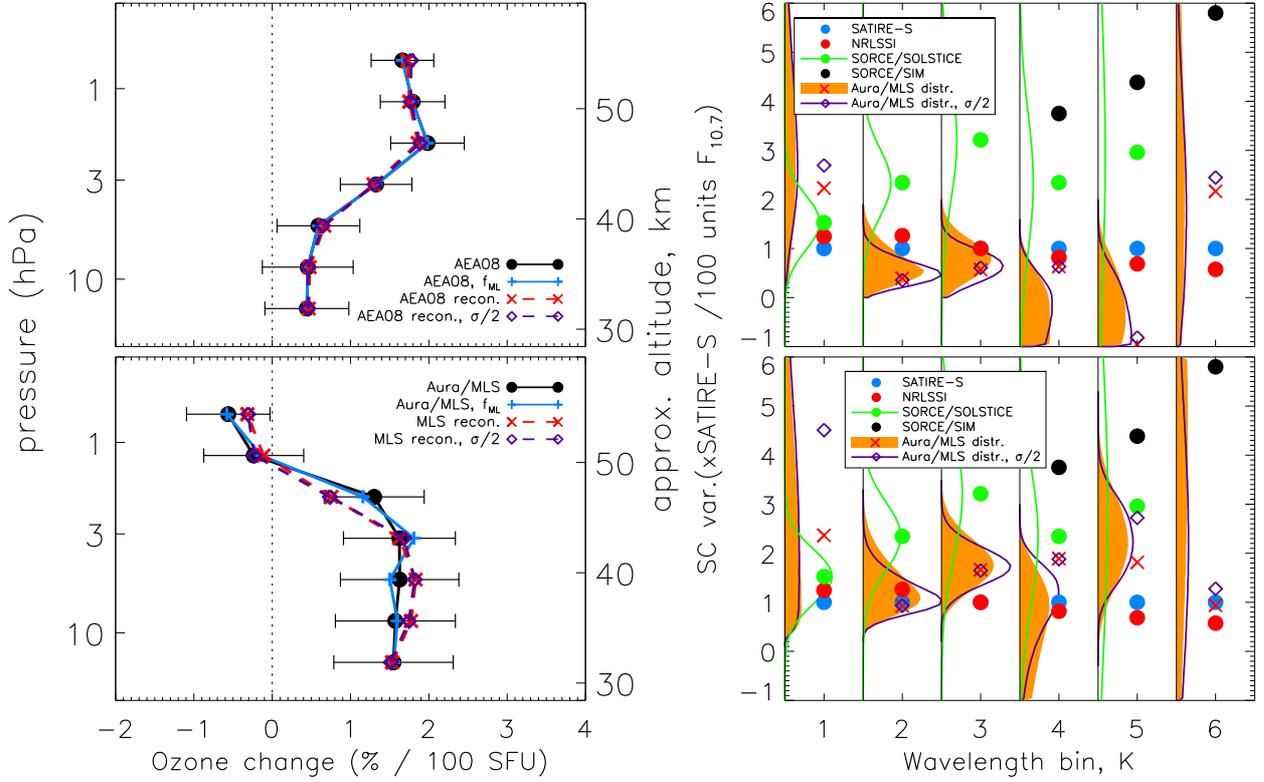}
\caption{The (left) $\Delta O_3(z)$ profile and the (right) respective SC SSI change in multiples of SATIRE-S for the six wavelength bands for (top) AEA08 and (bottom) Aura/MLS. The relative variability of SATIRE-S (blue), NRLSSI (red), SORCE/SOLSTICE (green) and SORCE/SIM (black) are represented by the filled circles. The green distribution is the uncertainty on the SORCE/SOLSTICE data. Posterior distributions are shown for the sampled SC changes using original (orange, filled) and halved (purple line) error bars. Red (crosses) and purple (diamonds) $\Delta O_3(z)$ profiles are the linear model profiles from the best-fit sampled parameters (also crosses and diamonds in the right-hand plots) with priors employed for the original and halved errors; the values of these individual, best-fit models are listed in columns~(4) and (6) of Table~\ref{fvaltable}.}
\label{figmls}
\end{figure*}

We calculated the posterior distribution of SC changes for four cases; results are presented in Figure~\ref{figmls} and Table~\ref{fvaltable}. The upper (lower) panels in Figure~\ref{figmls} are results using the AEA08 (Aura/MLS) ozone profile. The marginalised distribution of $f_k$ in each band, $\mathrm{Pr}(f_k|\textbf{d})$, is shown by the filled orange distributions in the right panels and the maximum likelihood (i.e., best fitting) sample from these posteriors are given in columns 4 and 6 of Table~\ref{fvaltable} and plotted as red crosses in Fig.~\ref{figmls}.

Despite such large differences between the mean profile shape of the AEA08 and Aura/MLS $\Delta O_3(z)$ profiles, the combination of data and prior assumptions, suggests that the most probable \textbf{f} parameters are closer to the SSI models than to SOLSTICE, except for band $k$ = 1 in both cases and in band $k=5$ in the Aura/MLS case. In general, to produce the Aura/MLS $\Delta O_3(z)$ profile, SC SSI changes need to be larger at all wavelengths than for AEA08.

If we ignore the priors set out in Section~\ref{priors} and allow any SC changes, however unphysical, to achieve the best fit to the $\Delta O_3(z)$ profiles (i.e., statistical inversion without any priors), then we get the maximum likelihood parameters, \textbf{f}$_{\mathrm{ML}}$ (columns~3 and 5 of Table~\ref{fvaltable}). The maximum likelihood parameters almost exactly reproduce the AEA08 $\Delta O_3(z)$ profile and is a close fit to the Aura/MLS profile, as shown in the left plots of Figure~\ref{figmls} with blue curves. The values of \textbf{f}$_{\mathrm{ML}}$ are outside the prior range in both cases, suggesting that the exact mean profile cannot represent $\Delta O_3(z)$ SC changes from the Sun alone. Indeed, \textbf{f}$_{\mathrm{ML}}$ for Aura/MLS exceed the prior range of SC flux changes, except for band $k=5$. These flux changes are so implausible that this signifies that a direct inversion of the mean observed ozone profile -- or trying to assess which SSI dataset best represents real SC flux changes in this way alone -- would lead to incorrect conclusions about SSI changes. The AEA08 profile inversion yields more plausible values of \textbf{f}$_{\mathrm{ML}}$, though bands $k=2$, 5 and 6 lie outside the prior range and, again, using this single result would lead to incorrect conclusions about SSI.

Now, if we include the large $\Delta O_3(z)$ profile uncertainties and apply the priors, we are able to reproduce similar $\Delta O_3(z)$ profiles using an \textbf{f} consistent with our understanding of SC changes from the best fit of the 10$^6$ values of \textbf{f} sampled from $\mathrm{Pr}(f_k|\textbf{d})$. This best fit is shown as red crosses in the right plots of Figure~\ref{figmls} with the corresponding $\Delta O_3(z)$ shown in the left plots by the red dashed curves. In both cases the fit is in good agreement with the observed profiles. The values, \textbf{f}, of the best fit are, in some cases, very different to the peak (mode) of the posterior distribution (e.g. bands $k=1$, 4 and 6 for Aura/MLS and $k=4$ for AEA08). The value of $f$ for which the posterior is peaked is the most probable a posteriori (MAP) model. This is distinct from the model defined by the peaks of the marginalised distributions in each of the six parameters separately (although this model also happens to fit the data reasonably well). Unless there is a particular reason to focus on a single wavelength band, it is the MAP model that should be considered.

\begin{figure*}
\noindent\includegraphics[width=40pc, bb=72 365 540 700]{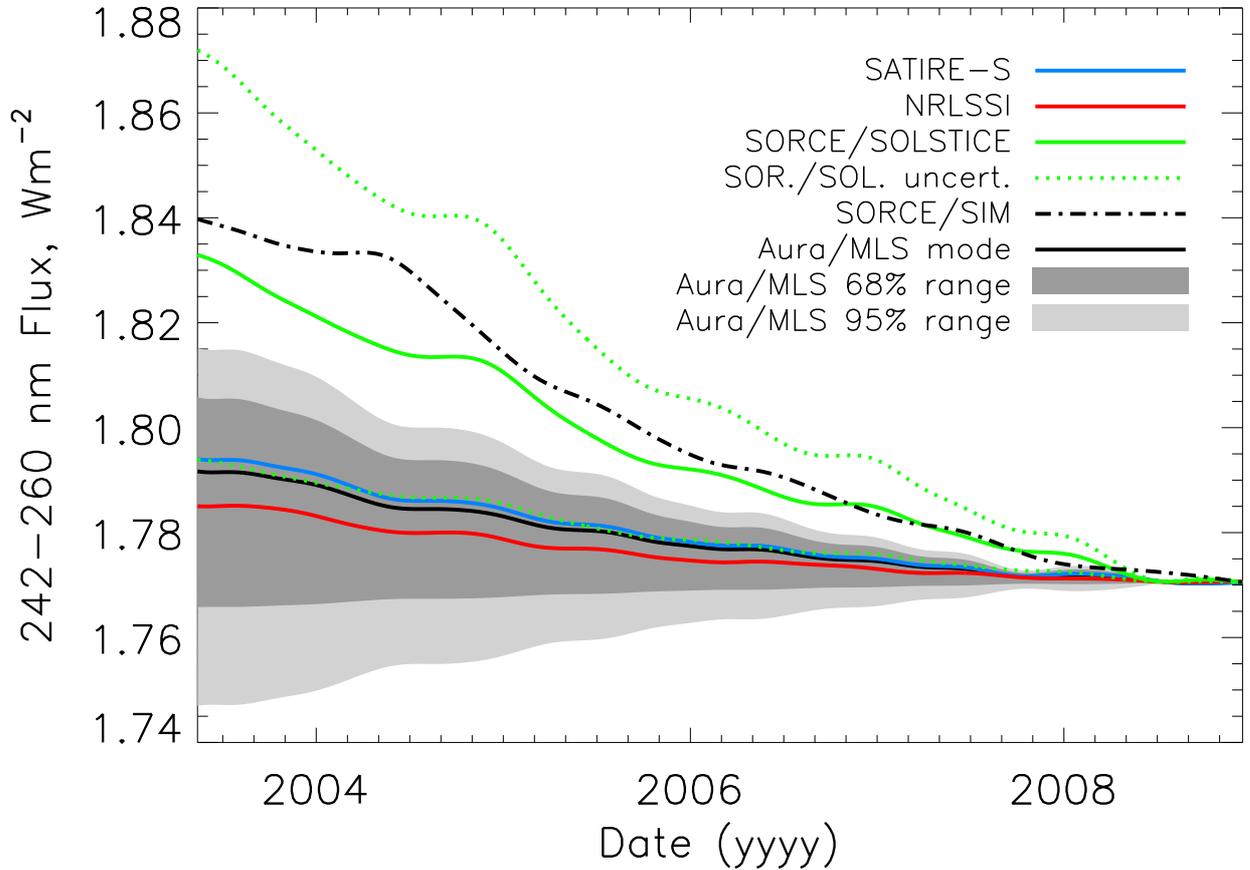}
\caption{The Aura/MLS posterior distribution at 242--260 nm, determined using the statistics presented in Fig.~\ref{figmls}, plotted as a timeseries for cycle 23, between 2003 and 2008. The mode is the solid black line, with the 68\% range in dark grey and 95\% range in light grey. Also shown is SATIRE-S (blue), NRLSSI (red), SORCE/SOLSTICE (green, solid line) with the SORCE/SOLSTICE uncertainty range (given by the dotted lines) and SORCE/SIM (black, dot-dashed line). The absolute values of all datasets have been adjusted by adding the difference at the solar minimum in December 2008 so all datasets have the same value at this time. All timeseries have been smoothed.}
\label{figssiexample}
\end{figure*}

An example of how these posterior distributions would appear as timeseries is given for the 242-260 nm band in Fig.~\ref{figssiexample}. The period shown covers SORCE/SOLSTICE observations beginning in May 2003 through to the end of SC 23, in December 2008; all datasets have been smoothed using a Gaussian window with an equivalent boxcar width of 135 days. The black line represents the mode result of Aura/MLS, with the 68\% range given in dark grey and the 95\% in light grey. For comparison, also included in the plot are the timeseries for SATIRE-S (blue), NRLSSI (red), SORCE/SOLSTICE (green, with uncertainty range given by dotted lines) and SORCE/SIM (black, dot-dashed line). The absolute values of all timeseries have been shifted to SATIRE-S by adding the mean difference in flux over the three month period centred at the solar minimum of December 2008.

The very wide spread of uncertainty associated with our result would be reduced given better knowledge of the ozone solar response and/or stronger priors. For example, repeating the analysis assuming the uncertainties in the ozone measurements to be halved has little effect on the reconstructed profiles (purple dashed curves in left hand panel of Fig.~\ref{figmls}), but does reduce the range of $\mathbf{f}$ (purple lines in the right-hand panel) in most bands. This shows that, given ozone data with smaller observational errors, our approach would be better able to constrain the possible range of SSI, though the uncertainties of the ozone profiles given here depend crucially on the accuracy of the other terms in the multiple linear regression analysis (see Section~\ref{eqprofiles}) that produce the SC response ozone profile. For example, with the ozone errors halved the analysis would imply that the SORCE/SOLSTICE values in bands $k=2$--5 are not consistent with the ozone SC change derived from the AEA08 data or $k=2$, 3 and 4 from Aura/MLS data; SORCE/SIM would not be consistent with both AEA08 and Aura/MLS. Other experiments (not shown) using more restrictive priors show much lower spread in the posteriors, but these still overlap largely with the distributions presented in Fig.~\ref{figmls}.

Finally, the red and purple profiles, as fits to the Aura/MLS profile in the lower left plot of Figure~\ref{figmls}, show clearly that different changes in SSI produce very similar $\Delta O_3(z)$ profiles. They are produced using very different values of $f_1$ and $f_5$, as given in column~4 of Table~\ref{fvaltable}. In other words, more than one combination (indeed, many different combinations) of the set of linear ozone profiles given in the \textbf{M}-matrix can produce almost identical ozone profiles. Thus, a simple comparison of modelled and observed SC changes (in ozone profile) cannot provide the necessary information to select the most likely one from a number of SSI datasets. To be able to distinguish between them is the ultimate goal of applying this method. While this is not currently achievable, the framework developed and the results presented in this work are the first steps in realising this aim.

\section{Conclusions}
\label{discuss}
We have investigated the relationship between ozone and SSI changes using Bayesian inference to incorporate both the uncertainty in ozone and prior knowledge about SSI variation.  Aside from having a rigorous basis in probability theory, Bayesian inference is particularly well suited to situations such as this in which quite distinct information, from different sources, is being utilised. As such, Bayesian inference should be applicable to a variety of solar and atmospheric problems in which no single data-set can provide a definitive result, and we are actively pursuing this line of investigation. Applied to the problem at hand, this approach shows that, because similar ozone profiles can be produced from different SC SSI changes, the current data are insufficient to distinguish between SSI models and observations. 

The method is developed on the basis, which we establish, that the ozone response to changes in SSI in a finite spectral band, at least in the tropical upper-stratosphere/lower-mesosphere, is close to linear and, furthermore, that the total change in ozone is simply the sum of that resulting from changes in the individual bands. 

We emphasise that the results presented here are based upon test cases designed to demonstrate our method and that we place no great store by the results, which depend on the validity of the observed profiles and their uncertainties. The two cases we provide here are based on the apparent ozone response to changes in SSI over 1979--2003 found by \cite{AustinTourpali2008} and in 2004--2012 in a new analysis of Aura/MLS data. In the future, the use of extended and/or improved observational ozone data should allow more robust estimates of the solar cycle flux changes. For example the SBUV V8.6 ozone data \citep{McPetersBhartia2013} provides a longer data record (1979-2012, and continuing to present) than the dataset used by \cite{AustinTourpali2008}, which combines the results from \cite{SoukharevHood2006}. 

The uncertainties in the SC $\Delta O_3(z)$ profiles are currently large and we have shown how reducing the size of these errors would help to constrain the range of implied SC SSI changes. A primary concern is the use of MLR to derive the SC $\Delta O_3(z)$ profiles. This assumes that the temporal variation in F10.7 cm solar flux behaviour is the same as that of solar UV irradiance in all wavelength bands that physically affect O$_{2}$/O$_{3}$ photolysis. It also assumes that the other proxies in the MLR analysis, e.g., QBO and volcanic aerosol, fully represent their variability, that other effects missing from the analysis do not influence the derived solar signal and that the time period of the data is representative of the behavior more generally. The two very different $\Delta O_3(z)$ profiles suggest that these concerns will not be addressed without longer-term measurements.

In the 242-260 nm band, shown in Fig.~\ref{figssiexample}, the SATIRE-S curve lies closest to the median line from our analysis and the NRLSSI curve is also well within the 68\% uncertainty range. The SORCE/SOLSTICE curve lies outside our 95\% range (and SORCE/SIM even further outside) but there is considerable overlap between the distributions of uncertainty. Overall, the SSI SC changes estimated for the two test cases appear more consistent with the modelled SSI dataset than the SORCE/SOLSTICE observations, but we can not conclude at this stage that these results can endorse one SSI dataset over another, neither modelled nor observed. This is because the results are based on assumptions that require further validation. Indeed, based on the assumptions that have gone into this analysis, it is impossible to state that any of SORCE/SOLSTICE, NRLSSI or SATIRE-S provides a more or less realistic reflections of the true solar cycle flux changes.

Despite the remaining problems, the approach presented in this paper is more robust than single profile comparisons and offers a significant advance in analytical techniques in the future. In the future we will be expanding and improving on this technique by (i) the inclusion of additional observational data, in the form of other O$_{3}$ datasets, the temperature response and other stratospheric constitutes, possibly over a greater \textit{z} range and using a 3D model, (ii) using the atmospheric and solar data synergistically, avoiding the necessity of making \textit{a priori} MLR estimates of the solar signal in the former and (iii) the use of stronger priors that include knowledge of the coherence in the solar spectrum (e.g. \citeauthor{CessateurDudokdeWit2011}, \citeyear{CessateurDudokdeWit2011}) that will further constrain the possible solar cycle variability. Indeed, given the large uncertainties on ozone observations, improvement (iii) is paramount to constraining the solutions and reaching stronger conclusions. By these means, we will further our understanding of solar UV variability and its effects on the middle atmosphere taking account of, as far as possible, all available datasets and their uncertainties.

\begin{acknowledgements}
We thank Peter Pilewskie and Marty Snow for helpful comments. We also thank the referees for their useful comments and suggestions that have led to an improved paper. For the multi-regression analysis on Aura/MLS data, we use data provided by KNMI Climate Explorer (climexp.knmi.nl) including the PMOD composite data from PMOD/WRC, Davos, Switzerland, sunspot data from the SIDC-team and Aura/MLS data from NASA JPL and GES DISC.
\end{acknowledgements}



\end{document}